\documentclass{elsart}
\usepackage{graphicx}
\usepackage{amsmath}
\usepackage{amssymb}
\def\Vec#1{\mbox{\boldmath $#1$}}
\input epsf.sty

\def\NPA{{Nucl. Phys.} {\bf A}}

\usepackage{amssymb}

\begin{document}
\begin{frontmatter}
\title
{Numerical Calculation of a Minimal Surface Using
Bilinear Interpolations and \\
Chebyshev Polynomials.}
\author
{Sadataka Furui$^1$ and Bilal Masud$^2$} 
\address
{$^1$School of Science and Engineering, Teikyo University,\\ Utsunomiya 320-8551, Japan\\
$^2$Center for High Energy Physics, Punjab University,\\
Lahore-54590, Pakistan
}
\ead{furui@umb.teikyo-u.ac.jp$^1$, bilalmasud@chep.pu.edu.pk$^2$}

\begin{abstract}
We calculate the minimal surface bounded by four-sided figures
whose projection on a plane is a rectangle, starting with the
bilinear interpolation and using, for smoothness, the Chebyshev
polynomial expansion in our discretized numerical algorithm to get
closer to satisfying the zero mean curvature condition. We report
values for both the bilinear and improved areas, suggesting a
quantitative evaluation of the bilinear interpolation.
An analytical expression of the Schwarz minimal surface with polygonal boundaries
and its 3-dimensional plot is also given.
\end{abstract}

\end{frontmatter}

\section{Introduction}
In mathematical modeling it is not uncommon to need a surface
that spans a known boundary and has the least value of a related
quantity, say, area. If the least area is desired, the problem is
termed in the mathematical literature as the \emph{Plateau
problem}, namely minimizing the area functional

\begin{equation}
\label{area def 1}
 A(X)=\int \int_{\Omega }|X_{u}\times X_{v}|\,du\,dv.
\end{equation}

Here $\Omega \subset R^{2}$ is a domain over which the surface $X$
is defined as a map, with the boundary condition $X(\partial
\Omega )=\Gamma $. It is known \cite{docarmo} that the first
variation of $A(X)$ vanishes if and only if the mean curvature $H$
of $X$ is zero \emph{everywhere} in it. Thus to get a minimal (or,
more precisely, a stationary) surface, we have to solve the
differential equation obtained by setting the mean curvature $H$
equal to zero \emph{for each value} of the two parameters, say,
$u$ and $v$ parameterizing a surface spanning the fixed boundary.
In a numerical work, the problem has to be discretized by choosing
a selection of the \emph{numerical} values of the two parameters
and finding the minimal- surface-position for each pair of the
values. If the given boundary is a four-sided figure whose
projection on a plane is a rectangle, the surface positions
become simply the heights above the $uv$%
-plane. Such `numerical heights' and resulting `numerical minimal
surfaces' have been computed in ref. \cite{businger} for a variety
of closed curve boundaries. 

In this paper, we report, in the
section \ref{numerical} below, a modification to their algorithm
that uses linear combinations of
the \emph{Chebyshev polynomials} as heights at the discretized $uv$%
-positions. In this way, we replace in the algorithm arbitrary
heights by linear combinations of convenient polynomials with
arbitrary coefficients.

The immediate advantage of this use of polynomials has been a
reduction in the discretization error and a better convergence.
Polynomials are (smooth) analytic functions having simply
calculable derivatives. We have also carried further our efforts
to find analytic surfaces that can be taken as `approximate
minimal surfaces':

 1) We read initial heights from a ruled analytic
surface spanning our fixed boundary, namely the \emph{bilinear
interpolation} introduced in the section \ref{bilinear}. And for
knowing how much heights changed through our numerical
minimization 

2) we compared the areas of the numerically found
points (or `numerical minimal surface' explained in section
\ref{area} with those of the bilinear interpolation for each of
the selected boundaries. 

Through this \emph{quantitative}
comparison, something missing in the previous works, we suggest to
a \emph{user} of a minimal surface bounded by four straight lines
a prescription that may well save almost all the computer
programming and CPU time spent in implementation, say, the
algorithm of refs. \cite{businger}: 
the approximate equality of the areas of the bilinear interpolation
and numerical minimal surface strongly suggests that the simple
bilinear interpolation \emph{itself} may work as a `minimal
surface' for many mathematical models that need minimal surfaces
bounded by four straight lines.

The only ruled surface, other than the
plane, which is a minimal surface is a helicoid \cite{docarmo}. As
one boundary of a helicoid must be part of a helix, which is not a
straight line, the boundary of a helicoid cannot be composed of
four straight lines. In this way there cannot be at least a ruled
surface which is a minimal surface bounded by four straight lines.

Since the calculation of the area given by the surface coordinates would be possible only numerically, it is technically important how to evaluate the 
minimal surface area accurately, and evaluate deviation from the ruled surface whose area can be evaluated analytically.
An area of bilinear interpolation is to be compared only with the numerically
calculated `minimal surfaces'.  (See the section
\ref{area} below for a description of the algorithms we used to
calculate areas of the `numerical minimal surfaces' along with the
resulting numerical area values.)

\section{Plateau problem}

 For a locally parameterized surface $\mathbf{X=X}(x,y,z(x,y))$, the mean
curvature $H$ is defined as

\begin{equation}
H =\frac{g_{11}h_{22}-2g_{12}h_{12}+g_{22}h_{11}}{%
g_{11}g_{22}-g_{12}^{2}},  \label{k def}
\end{equation}

\noindent where

\begin{equation}
g_{11}=\langle {\Vec X}_{u},{%
\Vec
X}_{u}\rangle, \hspace{.5 in} g_{12}=\langle {\Vec X}_{u},{%
\Vec X}_{v}\rangle \hspace{.5 in} {\rm and} \hspace{.5
in} g_{22}=
\langle {\Vec X}%
_{v},{\Vec X}_{v}\rangle
\end{equation}

\noindent are the 1st fundamental form and
\begin{equation}
h_{11}=\langle N,{\Vec X}_{uu}\rangle, \hspace{.5 in} h_{12}=\langle N,{%
\Vec X}_{uv}\rangle \hspace{.5 in} {\rm and} \hspace{.5 in} h_{22}=\langle N,{%
\Vec X}_{vv}\rangle
\end{equation}

\noindent are the second fundamental form. Here

\begin{equation}
 \displaystyle N=%
\frac{{\Vec
X}_{u}\times {\Vec X}_{v}}{|{\Vec X}%
_{u}\times {\Vec X}_{v}|}
\end{equation}

\noindent is the unit normal of the surface.

The vanishing condition of the numerator of $H$ becomes
\begin{equation}
F(z)=\frac{\partial ^{2}z}{\partial y^{2}}\left( 1+\left( \frac{\partial z%
}{\partial x}\right) ^{2}\right) -2\frac{\partial z}{\partial x}\frac{%
\partial z}{\partial y}\frac{\partial ^{2}z}{\partial x\partial y}+\frac{%
\partial ^{2}z}{\partial x^{2}}\left( 1+\left( \frac{\partial z}{\partial y}%
\right) ^{2}\right) =0  \label{Fz}
\end{equation}%

We are interested in evaluating the area bounded by skew quadrilateral\cite{FGM} whose boundary is composed
of four \emph{non-planar} straight lines connecting four corners $
\mathbf{x}_{00},\mathbf{x}_{01},\mathbf{x}_{10} \hspace{.1 in}
\rm{and} \hspace{.1 in} \mathbf{x}_{11}$. 

The Plateau problem for polygonal boundaries was studied by Schwarz, Weierstrass and Riemann \cite{Osserman,Schwarz,wrep}.

The minimal surface whose bounding contour is the skew quadrilateral consisting of four edges  
 $A(\frac{1}{2},0,\frac{1}{2\sqrt 2})$, $B(0,-\frac{1}{2},-\frac{1}{2\sqrt 2})$,
$C(-\frac{1}{2},0,\frac{1}{2\sqrt 2})$ and $D(0,\frac{1}{2},-\frac{1}{2\sqrt 2})$ was calculated by Schwarz\cite{Schwarz} using the Weierstrass-Enneper representation. An extensive derivation of the minimal surface is given in \cite{Osserman,Nitche}.  

In this theory, every simply connected, open minimal surface with normal domain  $\Pi$ is shown to be expressed in the form
\begin{equation}
\Vec r=\Vec r(\alpha,\beta)=\Vec r_0+Re\int_0^\gamma \Vec F(\gamma)d\gamma; \gamma\subset \Pi\
\end{equation}
where $\Vec F(\gamma)$ is a non-vanishing analytic vector in $\Pi$ satisfying ${\Vec F}^2=\phi_1^2(\gamma)+\phi_2^2(\gamma)+\phi_3^2(\gamma)=0$

One works with $\Phi(\gamma)=\sqrt{(\phi_1(\gamma)-i\phi_2(\gamma))/2}$ and
$\Psi(\gamma)=\sqrt{(\phi_1(\gamma)+i\phi_2(\gamma))/2}$ and $2\Phi\Psi=\phi_3$

When $\Phi$ and $\Psi$ do not have the common zero, the following expression was obtained:
\begin{eqnarray}
x&=&x_0+Re \int_0^\gamma (\Phi^2-\Psi^2) d\gamma\nonumber\\
y&=&y_0+Re \int_0^\gamma i(\Phi^2+\Psi^2) d\gamma\nonumber\\
z&=&z_0+Re \int_0^\gamma 2\Phi \Psi d\gamma
\end{eqnarray}

Using the mapping $\omega(\gamma)=\Psi(\gamma)/\Phi(\gamma)$, and defining $\Phi(\gamma)^2d\gamma=R(\omega)d\omega$, Schwarz obtained the expression
\begin{eqnarray}\label{schwarz}
x&=&Re \int^\omega (1-\omega^2)R(\omega) d\omega\nonumber\\
y&=&-Im \int^\omega (1+\omega^2)R(\omega) d\omega\nonumber\\
z&=&Re \int^\omega 2\omega R(\omega) d\omega
\end{eqnarray}
where 
\[
R(\omega)=-\frac{2}{\sqrt{1+14\omega^4+\omega^8}}
\]

The integral can be done analytically, whose detail is given in the Appendix.

\section{The Bilinear Interpolation:}
\label{bilinear}

We try to approach the minimal surface for the boundary composed
of four \emph{non-planar} straight lines connecting four corners $
\mathbf{x}_{00},\mathbf{x}_{01},\mathbf{x}_{10} \hspace{.1 in}
\rm{and} \hspace{.1 in} \mathbf{x}_{11}$ by improving upon a
surface that spans this boundary, namely a \emph{hyperbolic
paraboloid} \cite{farin}

\begin{equation}
\mathbf{x}(u,v)=[1-u \ \ \ \ \ \ u]\left[
\begin{array}{cc}
\mathbf{x}_{00} & \mathbf{x}_{01} \\
\mathbf{x}_{10} & \mathbf{x}_{11}
\end{array}
\right] \left[
\begin{array}{c}
1-v \\
v \end{array} \right] \label{hyper parabola}
\end{equation}

\noindent (Hyperbolic paraboloid is a bilinear interpolation; it
might interest the reader that this is a special case of the
general bilinear interpolation, termed the \emph{Coons Patch
}\cite{farin}.) For the corners we chose, for a selection of
integer values of $d$ and $r$:

\begin{equation}
\mathbf{x}_{00}=\mathbf{r}_{1}\hspace{0.5in}\mathbf{x}_{10}=\mathbf{r}_{4}
\hspace{0.5in}\mathbf{x}_{01}=\mathbf{r}_{3}\hspace{0.5in}\mathbf{x}_{11}=
\mathbf{r}_{2}
\end{equation}

We consider two types of configurations of the four corners: ruled$_1$ and ruled$_2$.  In the case of ruled$_1$ we choose
\begin{equation}
\mathbf{r}_{1}=(0,0,0)\hspace{0.4in}\mathbf{r}
_{2}=(r,d,0)\hspace{0.4in}\mathbf{r}_{3}=(0,d,d)\hspace{0.4in}\mathbf{r}
_{4}=(r,0,d).
  \label{r1234_1}
\end{equation}
The mapping from $(u,v)$ to $(x,y,z)$ in this case is
\begin{eqnarray}
x(u,v)&=&u r\nonumber\\
y(u,v)&=&v r\nonumber\\
z(u,v)&=&u d+v d(1-2u)
\end{eqnarray}

In the case of ruled$_2$ we choose
\begin{equation}
\mathbf{r}_{1}=(0,0,0)\hspace{0.4in}\mathbf{r}
_{2}=(r,r,0)\hspace{0.4in}\mathbf{r}_{3}=(0,r,d)\hspace{0.4in}\mathbf{r}
_{4}=(r,0,d).
  \label{r1234_2}
\end{equation}
The mapping from $(u,v)$ to $(x,y,z)$ in this case is
\begin{eqnarray}
x(u,v)&=&u r\nonumber\\
y(u,v)&=&v d\nonumber\\
z(u,v)&=&u d+v d(1-2u)
\end{eqnarray}

\noindent These definitions are such that for $r=d$ the four
position vectors lie at the corners of a regular tetrahedron. The
Fig. 1 and Fig.2 below are 3D graphs of the hyperbolic paraboloid for a
choice of corners mentioned in eqs.(\ref{r1234_1})and (\ref{r1234_2}).

\begin{figure}
       \centerline{\includegraphics{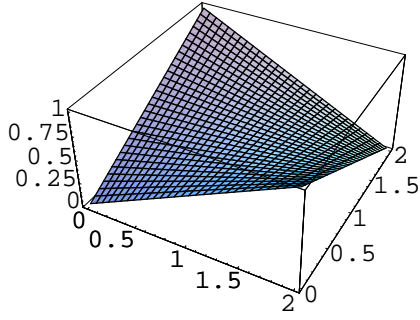}}
       \caption{The ruled$_1$ surface (r=1,d=2). The horizontal plane is expanded by $y,z$ and the height is $x$.}
        \label{fig:1}
\end{figure}
\begin{figure}
       \centerline{\includegraphics{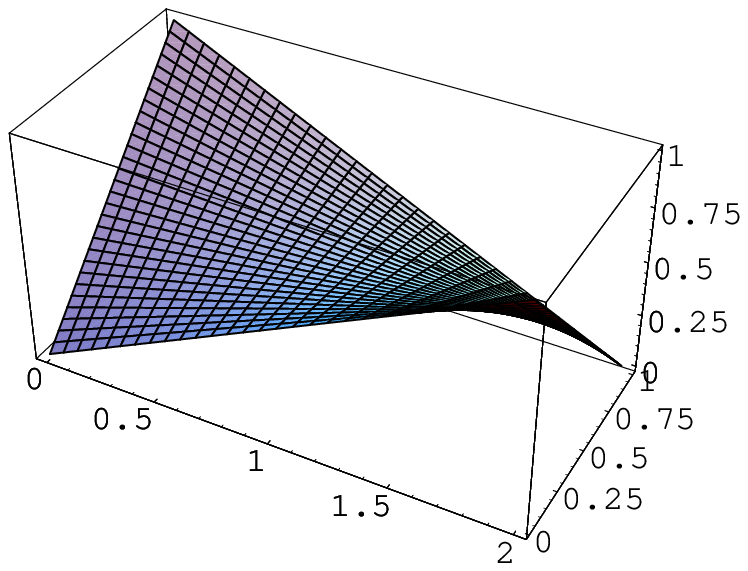}}
       \caption{The ruled$_2$ surface (r=2,d=1). The horizontal plane is expanded by $x,y$ and the height is $z$.}
        \label{fig:2}
\end{figure}

\begin{figure}
       \centerline{\includegraphics{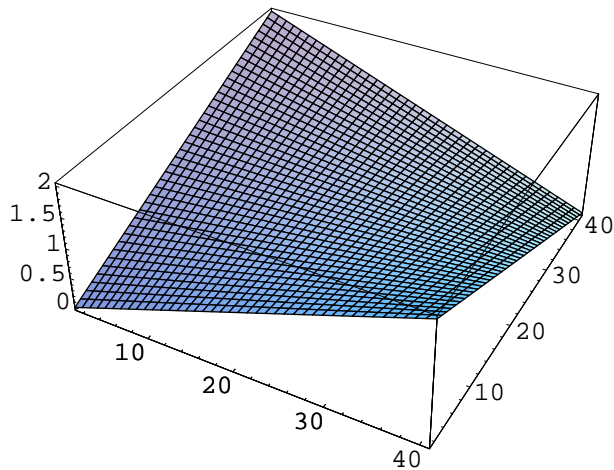}}
       \caption{The numerically fitted ruled$_1$ surface (r=1,d=2).}
        \label{fig:3}
\end{figure}

For a surface to be minimal, its mean curvature vanishes
everywhere \cite{docarmo}. The expression for the mean curvature,
calculated using eq.(\ref{k def}) of our bilinear interpolation is

\begin{equation}
\frac{-2d r(1-2u)(1-2v)}{%
[r^{2}+r^{2}(1-2u)^2+d^{2}(1-2v)^2]^{3/2}}. \label{m curv_1}
\end{equation}
for the ruled$_1$ and
\begin{equation}
\frac{-2r^{3}(1-2u)(1-2v)}{%
d[d^{2}+r^{2}(1-2u)^2+r^{2}(1-2v)^2]^{3/2}}. \label{m curv_2}
\end{equation}
for the ruled$_2$.

\noindent The mean curvature for the surface is zero only for the $u=\frac{1%
}{2}$ line and the $v=\frac{1}{2}$ line, whereas for a minimal
surface this should be zero for all values of $u$ and $v$.

\section{The Numerical Work:}
\label{numerical}
The solution of the Plateau problem was formulated by Courant\cite{Courant} as  minimization of the Dirichlet integral
\[
E_D(u)=\frac{1}{2}\int_\Omega |\nabla u|^2
\]
where $|\nabla u|^2={\rm tr} (^t \partial u \partial u)$, where $\partial u$ is the matrix of partial derivatives of u in an orthonormal basis. In \cite{pipo}, a mapping to the conjugate minimal surface was considered in the minimization process.
In \cite{dziuk}, a diffeomorphism $u_0:\Omega\to {\bf R}^3$, where $\Omega\subset {\bf R}^2$, $u_0(\Omega)\subset S(0)$ and maps $u(\cdot, t):\Omega\to {\bf R}^3, (t>0)$, which satisfies
\[
\frac{\partial u}{\partial t}-\Delta_{S(t)}u=0 \quad {\rm in} \Omega\times(0,T)
\]
with appropriate Dirichlet boundary condition was considered. 

In \cite{businger}, more direct minimization of the numerator of the mean curvature $H$ using parallel computer was performed. 
In the generalized Newton's method, the minimization of $F(z)$ of
eq.(\ref{Fz}) is achieved by the iteration
\begin{equation}
z^{k+1}=z^{k}-DF(z^{k})^{-1}[F(z^{k})],
\end{equation}
\noindent where $DF(z^{k})^{-1}$ is the inverse of the functional
derivative that satisfies
\begin{equation}
DF(z^{k})[z^{k+1}-z^{k}]=-F(z^{k}).
\end{equation}

 We consider $(N+1)\times (N+1)$ lattice grid points $(u_{i},v_{j})$%
, $(0\leq i\leq N,0\leq j\leq N)$ and corresponding
$z(u_{i},v_{j})$. We keep same number of grid points independent of $r$ and $d$.
In the discretized system $z^{k+1}(u_i,v_j)$ is defined from
$z^k(u_i,v_j)$ by adding $dz^{k+1}(u_i,v_j)$ which can be
calculated by solving the linear equation expressed by a matrix
$C$ defined by the first and the second fundamental form as
\begin{equation}
C adz^{k+1}(u,v)=-F(z^k(u,v))
\end{equation}

Businger et. al. \cite{businger} gave a Mathematica code to define
the matrix $C$. In our problem of improving the surface starting
from the bilinear area, the discretization error in the
replacement like
\begin{equation}
\frac{\partial z}{\partial
u}=\frac{z(u_{i+1},v_{j})-z(u_{i},v_{j})}{du}
\end{equation}%
\noindent is large and the convergence was poor.

The reason would be lack of explicit third order polynomial term
in the evaluation of $dz^{k+1}$ in the numerical methods which
manifests itself in the fact that $C_{(i-1,j)}$ and $C_{(i+1,j)}$
are identical. Thus we evaluate the first and second fundamental
form on the discretized system by using the Chebyshev
polynomial expansion \cite{NR}.

\subsection{Chebyshev Polynomial Expansion}

The Chebyshev polynomial of degree $n$ is denoted $T_{n}(x)$ and
is given by
\begin{equation}
T_{n}(x)=\cos (n\cos ^{-1}x)
\end{equation}%
where the range of $x$ is $[-1,1]$ and their explicit expressions
are given by the recursion
\begin{equation}
T_{0}(x)=1,\quad T_{n+1}(x)=2xT_{n}(x)-T_{n-1}(x)\quad (n\geq 1)
\end{equation}%
\noindent The zeros of $T_{n}(x)$ are located at
\begin{equation}
x_{k}=\cos (\frac{\pi (k+1/2)}{n}),\quad (k=0,1,\cdots ,n-1)
\end{equation}%
\noindent If $x_k$ $(k=0,1,\cdots,m-1)$ are the $m$ zeros of $T_m(x)$,  the Chebyshev polynomial satisfies the discrete
orthogonality relation for $i,j<m$,

\begin{equation}
\sum_{k=0}^{m-1}T_{i}(x_{k})T_{j}(x_{k})=\left\{
\begin{array}{ll}
0 & i\neq j \\
m/2 & i=j\neq 0 \\
m & i=j=0%
\end{array}%
\right.
\end{equation}

We first map $z(u_{i},v_{j}),(0\leq u_{i}\leq 1,0\leq v_{j}\leq 1)$ onto $%
z(x_{i},y_{j}),(-1\leq x_{i}\leq 1,-1\leq y_{j}\leq 1)$ and
interpolate values at zeros of the $T_{N+1}(x)$ defined as $x_{l}$
$(l=0,1,\cdots ,N)$
and $T_{N+1}(y
)$ defined as $y_{m}$, $(m=0,1,\cdots ,N)$, i.e. $%
z(x_{l},y_{m})$.

We define $c(x_{l},n)$ $(n=0,\cdots ,N)$ as%

\begin{equation}
c(x_{l},n)=\frac{2}{N+1}\sum_{m=1}^{N+1}z(x_{l},y_{m})T_{n}(y_{m})
\end{equation}%
\noindent and interpolate at $y=y_{j}$ via

\begin{equation}
\hat{z}(x_{l},y_{j})=\sum_{n}c(x_{l},n)T_{n}(y_{j})-\frac{1}{2}c(x_{l},0)
\end{equation}%
\noindent Partial derivative in $y$ is performed by replacing $T_{n}(y)$ by %

\begin{equation}
\displaystyle\frac{dT_{n}(y)}{dy}=T_{n}^{\prime }(y)
\partial _{y}\hat{z}(x_{l},y_{j})=\sum_{n}c(x_{l},n)T_{n}^{\prime }(y_{j}).
\end{equation}

So far the $x$-coordinate is restricted to zero points $x^{l}$.
Now, interpolation to $x=x_{i}$ is performed by

\begin{equation}
\tilde{c}(n,y_{j})=\frac{2}{N+1}\sum_{l=1}^{N+1}\hat{z}%
(x_{l},y_{j})T_{n}(x_{l})
\end{equation}%
\noindent We define also $\partial _{y}\tilde{c}(n,y_{j})$ as
\begin{equation}
\partial _{y}\tilde{c}(n,y_{j})=\frac{2}{N+1}\sum_{l=1}^{N+1}\partial _{y}%
\hat{z}(x_{l},y_{j})T_{n}^{\prime }(x_{l})
\end{equation}%
\noindent The values on the mesh points $\tilde{z}(x_{i},y_{j})$
are

\begin{equation}
\tilde{z}(x_{i},y_{j})=\sum_{n}\tilde{c}(n,y_{j})T_{n}(x_{i})-\frac{1}{2}%
\tilde{c}(0,y_{j})
\end{equation}
\noindent and the derivatives $\partial
_{x}\tilde{z}(x_{i},y_{j})$ and $\partial
_{x}^{2}\tilde{z}(x_{i},y_{j})$ are
\begin{equation}
\partial _{x}\tilde{z}(x_{i},y_{j})=\sum_{n}\tilde{c}(n,y_{j})T_{n}^{\prime
}(x_{i})  \label{zderiv}
\end{equation}

\begin{equation}
\partial _{x}^{2}\tilde{z}(x_{i},y_{j})=\sum_{n}\tilde{c}(n,y_{j})T_{n}^{%
\prime \prime }(x_{i})
\end{equation}

\begin{equation}
\partial _{x}\partial _{y}\tilde{z}(x_{i},y_{j})=\sum_{n}\partial _{y}\tilde{%
c}(n,y_{j})T_{n}^{\prime }(x_{i})
\end{equation}

 In the linear equation
\begin{equation}
{Cdz^{k+1}}_{(i,j)}=b_{(i,j)}
\end{equation}%
the matrix $C$ in the left-hand side(lhs) is a sparse matrix that
contains at least nine non-vanishing elements in each row. Around
the position ${(i,j)}$ $(0\leq i\leq N,0\leq j\leq N)$ the
elements for the nine nearest neighbors of $(i,j)$ are

\begin{eqnarray}
C_{(i-1,j-1)} &=&-\partial _{y}\tilde{z}(x_{i},y_{j})\cdot \partial _{x}%
\tilde{z}(x_{i},y_{j})/(2.\cdot du\cdot dv)  \nonumber \\
C_{(i-1,j)} &=&(1+\partial
_{y}\tilde{z}(x_{i},y_{j})^{2})/du^{2}-\partial
_{y}^{2}\tilde{z}(x_{i},y_{j})\cdot \partial
_{x}\tilde{z}(x_{i},y_{j})/du
\nonumber \\
C_{(i-1,j+1)} &=&\partial _{y}\tilde{z}(x_{i},y_{j})\cdot \partial _{x}\tilde{%
z}(x_{i},y_{j})/(2\cdot du\cdot dv)  \nonumber \\
C_{(i,j-1)} &=&(1+\partial
_{x}\tilde{z}(x_{i},y_{j})^{2})/dv^{2}+\partial
_{x}\tilde{z}(x_{i},y_{j})\cdot \partial _{x}\partial _{y}\tilde{z}%
(x_{i},y_{j})/dv\nonumber\\
&&-\partial _{y}\tilde{z}(x_{i},y_{j})\cdot \partial _{x}^{2}%
\tilde{z}(x_{i},y_{j})/dv  \nonumber \\
C_{(i,j)} &=&-2\cdot (1+\partial
_{y}\tilde{z}(x_{i},y_{j})^{2})/du^{2}-2\cdot
(1+\partial _{x}\tilde{z}(x_{i},y_{j})^{2})/dv^{2}  \nonumber \\
C_{(i,j+1)} &=&(1+\partial
_{x}\tilde{z}(x_{i},y_{j})^{2})/dv^{2}-\partial
_{x}\tilde{z}(x_{i},y_{j})\cdot \partial _{x}\partial _{y}\tilde{z}%
(x_{i},y_{j})/dv\nonumber\\
&&+\partial _{y}\tilde{z}(x_{i},y_{j})\cdot \partial _{x}^{2}%
\tilde{z}(x_{i},y_{j})/dv  \nonumber \\
C_{(i+1,j-1)} &=&\partial _{y}\tilde{z}(x_{i},y_{j})\cdot \partial _{x}\tilde{%
z}(x_{i},y_{j})/(2\cdot du\cdot dv)  \nonumber\\
C_{(i+1,j)} &=&(1+\partial
_{y}\tilde{z}(x_{i},y_{j})^{2})/du^{2}+\partial
_{y}^{2}\tilde{z}(x_{i},y_{j})\cdot \partial _{x}\tilde{z}(x_{i},y_{j})/du\nonumber\\
&&-\partial _{y}\tilde{z}(x_{i},y_{j})\cdot \partial _{x}\partial _{y}\tilde{z}
(x_{i},y_{j})/du\nonumber\\
C_{(i+1,j+1)} &=&-\partial _{y}\tilde{z}(x_{i},y_{j})\cdot \partial _{x}%
\tilde{z}(x_{i},y_{j})/(2\cdot du\cdot dv)
\end{eqnarray}

The right hand side is
\begin{eqnarray}
b_{(i,j)}&=&2\cdot\partial_x \tilde z(x_i,y_j)\cdot\partial_x\partial_y
\tilde z(x_i,y_j)\cdot\partial_y \tilde z(x_i,y_j)-\partial_x^2 \tilde
z(x_i,y_j)\cdot(1+\partial_y \tilde z(x_i,y_j)^2)\nonumber\\
&-&(1+\partial_x \tilde
z(x_i,y_j)^2)\cdot\partial_y^2 \tilde z(x_i,y_j)
\end{eqnarray}

\noindent The linear equation
\begin{equation}
{Cdz^{k+1}}_{(i,j)}=b_{(i,j)}
\end{equation}%
for $(N-1)\times (N-1)$ length's vector corresponding to the points inside the boundary can be solved by using standard computer library.

In the actual numerical calculation we multiply a reduction factor
to the solution $dz^{k+1}$ in each step to control the convergence.

\subsection{Evaluation of the Area}
\label{area}

The standard expression \cite{docarmo} in the differential
geometry for the area of a regular surface \textbf{x}$(u,v)$
parameterized in terms of two scalar parameters $u$ and $v$ is

\begin{equation}
\rm{area} =\int\limits_{{}}^{{}}du\int\limits_{{}}^{{}}dv\,|{\mathbf x}%
_{u}\times {\mathbf x}_{v}|,
  \label{area def}
\end{equation}

\noindent with

$\displaystyle {\mathbf x}_{u}\equiv \frac{\partial \mathbf x}{\partial u}$
and $\displaystyle{\mathbf x}_{v}\equiv \frac{\partial \mathbf x}{\partial v}.$ For $%
\mathbf{x=x}(x,y,z(x,y)),$ this becomes \cite{docarmo}

\begin{equation}
\rm{area} =\int_{Q}\sqrt{1+z_{x}^{2}+z_{y}^{2}}\,dx\,dy,
\end{equation}

\noindent where $Q$ is the normal projection of the surface onto
the $xy$ plane. Accordingly, we calculated the area formed by the
above mentioned discrete points as
\begin{equation}
\sum_{(i,j)}\sqrt{1+\partial _{x}\tilde{z}(x_{i},y_{j})^{2}+\partial _{y}%
\tilde{z}(x_{i},y_{j})^{2}}du\cdot dv.  \label{areac}
\end{equation}%
\noindent This expression contains discretization errors. To
estimate that, we discretized the bilinear interpolation for
$r=d=1$ in eq.(\ref{r1234_1}) as a $31\times 31$ grid, and
calculated the area obtained (of the discrete points) by this eq.
(\ref{areac}). This gave 1.2717 i.e. 0.7\% underestimation of the
exact value 1.280789 obtained by eq.(\ref{area def}).

\begin{figure}
       \centerline{\includegraphics{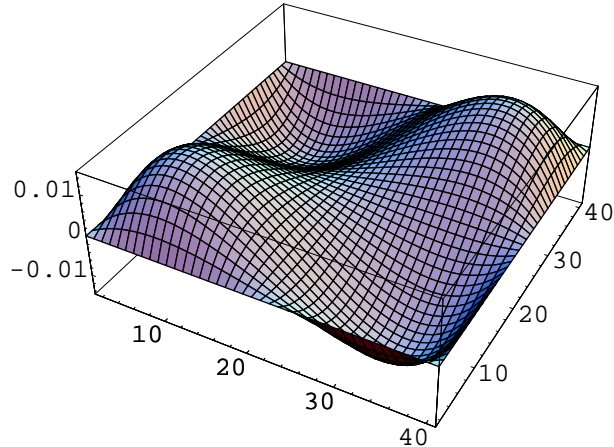}}
       \caption{The difference of the numerical minimum and the ruled surface (r=1,d=1).}
        \label{fig:4}
\end{figure}

\begin{figure}
       \centerline{\includegraphics{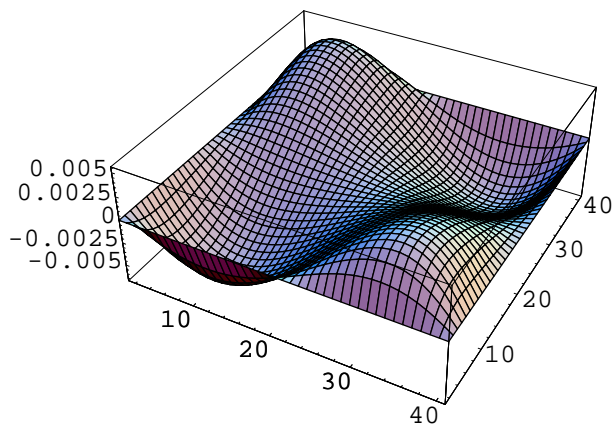}}
       \caption{The difference of the numerical minimum and the ruled surface (r=2,d=1).}
        \label{fig:5}
\end{figure}

In Fig.\ref{fig:4}, we show difference of the numerically calculated (N=40) minimal surface and the ruled$_1$ surface for $r=d=1$. The corresponding difference of $r=2,d=1$ is shown in Fig.\ref{fig:5}. 

That indicated that before reporting our `numerical areas' we
should compare different algorithms for calculating area out of a
given set of points. Thus, we calculated the area by the sum of
triangle $S_{1}$ spanned by

${\Vec
v}_{(i,j)}^{1}=(0,dv,z(u_{i},v_{j})-z(u_{i},v_{j-1}))$
and ${\Vec v}%
_{(i,j)}^{2}=(du,0,z(u_{i},v_{j})-z(u_{i-1},v_{j}))$, and
$S_{2}$
spanned by ${\Vec v}_{(i,j)}^{2}$ and ${\Vec v}%
_{(i,j)}^{3}=(du,dv,z(u_{i},v_{j})-z(u_{i-1},v_{j-1}))$

\begin{equation}  \label{areab}
\sum_{(i,j)}(|{\Vec v}^1_{(i,j)}\times {\Vec v}%
^2_{(i,j)}|+|{\Vec v}^2_{(i,j)}\times {\Vec v}%
^3_{(i,j)}|)/2
\end{equation}

\noindent The sum of triangles evaluated by the cross products is
1.281277037,i.e. 0.038\% overestimation.

The sum of triangles in the case of N=41 is 1.2811 i.e. 0.02\%
overestimation and in the case of N=21 is 1.2819 i.e. 0.09\%.

We also used a computer algebra system \cite{math} to find the
two-dimensional interpolation surface working by fitting
polynomial curves between successive data points followed by
finding areas of the analytical interpolation surface $\mathbf{x}$
by an exact double integral of eq.(\ref{area def}). The order 2
interpolation gave the above area as 1.280789195, the same up to 7 decimal places as the area without any discretization.

Guided by this check, for areas formed by points we report both
the areas calculated by triangulation as well by the
interpolation-followed-by-the-double-integral; the numerical
values strongly suggest these as better algorithms than the one
used in eq.(\ref{areac}).


The area of the ruled surface can be calculated analytically\cite{FGM}. In the Appendix, we give formulae of the area of the ruled$_1$ surface and the ruled$_2$ surface.  Numerically calculated area of the minimal surfaces( corresponding to the ruled$_2$ surface) and analytically calculated area of the ruled surfaces for given $r$ and $d$ are compared in Table.\ref{numeric}. The error bars are estimated from the convergence of the iteration. Numerical minimal surfaces corresponding to the ruled$_1$ are also slightly smaller than the analytical results.
In the numerical calculation, approach to the absolute minimum is not guaranteed. In a variational calculation we could obtain slightly smaller area.

\begin{table}[htb]
\begin{center}
\begin{tabular}{c|c|c|c}
$r,d$ & numerical area&  ruled$_2$ area & ruled$_1$ area  \\ \hline
1, 1 & 1.2793(5) &1.280789275& 1.280789275  \\ 
2, 1 & 2.3665(5)  & 2.366974371& 1.861564196 \\ 
1, 2 & 3.1753(5)  & 3.180414498 & 4.316148066\\ 
3, 1 & 3.4916(5)  & 3.491711893& 2.595828045 \\
1, 3 & 5.9310(5)   & 5.936348433 & 9.325179471\\
3, 2 & 7.2582(5)  & 7.259880701& 6.208799631\\ 
2, 3 & 8.5226(5) & 8.527786411 & 10.22064879 \\
\hline
\end{tabular}
\caption{ The numerical area  (calculated using the order 2
interpolation) and the analytical area of the ruled$_2$ surface for
the hyperbolic paraboloid of given $r$ and $d$. Analytical area of the ruled$_1$ surface is added for comparison.
}\label{numeric}
\end{center}
\end{table}

An explicit analytical calculation of the minimal surface in ${\bf R}^3$ is given in Appendix 2.
By constructing the conjugate minimal surface, Karcher\cite{Karcher} transformed the plateau problem in ${\bf R}^3$ into that in ${\bf S}^3$ and showed that the global Weierstrass representation of triply periodic minimal surfaces is possible.  We do not know whether the analytical calculation of the amount of the exact minimal surface area is possible through this method. 

We showed in the Appendix B that the exact minimal surface of Schwarz can be visualized. In order to evaluate the area, however, we need to interpolate the analytically obtained coordinates of the surface and perform numerial integration. We leave this task as a future study. 
Accurate numerical evaluation of the amount of the area is important for physical application and the Chebyschev polynomial expansion is a practical method for performing this process since the area is parametrized as $(x,y, z(x,y)$ instead of $(x(r,\theta),y(r,\theta),z(r,\theta))$. 

\section*{Acknowledgement}
 We thank the referee for drawing our attention to the analytical results of Schwarz reviewed in Ref.\cite{Nitche} and numerical approaches. S.F. thanks the Wolfram research staff Roger Germundsson for the information on the 3D graphics of "Mathematica" ver. 6.
The numerical calculation using the Chebyschev Polynomial was done by Hitachi SR8000 at High Energy Accelerator Research Organization (KEK).

\appendix
\section{Appendix 1: Area of the ruled surface}
In this Appendix, we present the analytical formulae of the area of ruled surfaces\cite{FGM}.
\subsection{Ruled$_1$ surface}
In the case of ruled$_1$ surface we define ${\Vec r}_{12}=(r,d,0)$, ${\Vec r}_{43}=(-r,d,0)$, ${\Vec r}_{23}=(-r,0,d)$, ${\Vec r}_{14}=(r,0,d)$.
The area is given by
\begin{equation}
\int_0^1du\int_0^1dv|[u{\Vec r}_{12}+(1-u){\Vec r}_{43}]\times[v{\Vec r}_{23}+(1-v){\Vec r}_{14}]|
\end{equation}
which becomes
\begin{eqnarray}
&&\int_0^1du\int_0^1 dv d\sqrt{d^2+2r^2(1-2u+2u^2)-4r^2v(1-v)}\nonumber\\
&&=d[\sqrt{d^2+2r^2}/3\nonumber\\
&&-2d^3 \tan^{-1}\left[\frac{dr(6d^4+2d^2r^2-(4d^2+r^2)r\sqrt{d^2+2r^2})}{-4d^6+4d^2r^4+r^6}\right]/(12r^2)\nonumber\\
&&+2d^3 \tan^{-1}\left[\frac{dr(6d^4+2d^2r^2+(4d^2+r^2)r\sqrt{d^2+2r^2})}{-4d^6+4d^2r^4+r^6}\right]/(12r^2)\nonumber\\
&&-\frac{3d^2+r^2}{6r}\log\left|\frac{-r+\sqrt{d^2+2r^2}}{r+\sqrt{d^2+2r^2}}\right|]\nonumber\\
&&=d\left[\sqrt{d^2+2r^2}/3+d^3 \tan^{-1}[\frac{2r^2 d\sqrt{d^2+2r^2}}{r^4-2r^2d^2-d^4}]/(6r^2)\right.\nonumber\\
&&\left.-\frac{3d^2+r^2}{6r}\log\left|\frac{-r+\sqrt{d^2+2r^2}}{r+\sqrt{d^2+2r^2}}\right|\right].
\end{eqnarray}

\subsection{Ruled$_2$ surface}
The ruled$_2$ surface is characterized by
${\Vec r}_{13}=(0,d,d)$, ${\Vec r}_{42}=(0,d,-d)$, ${\Vec r}_{23}=(-r,0,d)$, ${\Vec r}_{41}=(-r,0,-d)$.
The area is given by
\begin{equation}
\int_0^1du\int_0^1dv|[u{\Vec r}_{13}+(1-u){\Vec r}_{42}]\times[v{\Vec r}_{23}+(1-v){\Vec r}_{41}]|
\end{equation}
which becomes
\begin{eqnarray}
&&\int_0^1du\int_0^1 dv d\sqrt{d^2+2r^2(1-2u+2u^2)+d^2(1-2v)^2}\nonumber\\
&&=d\sqrt{d^2+2r^2}/3+\frac{r^2}{6} \tan^{-1}[\frac{d\sqrt{d^2+2r^2}}{r^2}]\nonumber\\
&&+\frac{r^2}{3}\log\left|\frac{d+\sqrt{d^2+2r^2}}{-d+\sqrt{d^2+2r^2}}\right|+\frac{dr}{4}(1+\frac{d^2}{3r^2})\log\left|\frac{r+\sqrt{d^2+2r^2}}{-r+\sqrt{d^2+2r^2}}\right|.
\end{eqnarray}

\section{Appendix 2: Visualization of the exact minimal surface}
In this Appendix, we construct conformal mapping from a complex $\omega$ plane to the skew quadrilateral of Schwarz, and visualize the surface using Mathematica\cite{math}. 

\begin{figure}[hb]
\begin{minipage}[b]{0.47\linewidth}
\begin{center}
\includegraphics[width=5cm,angle=0,clip]{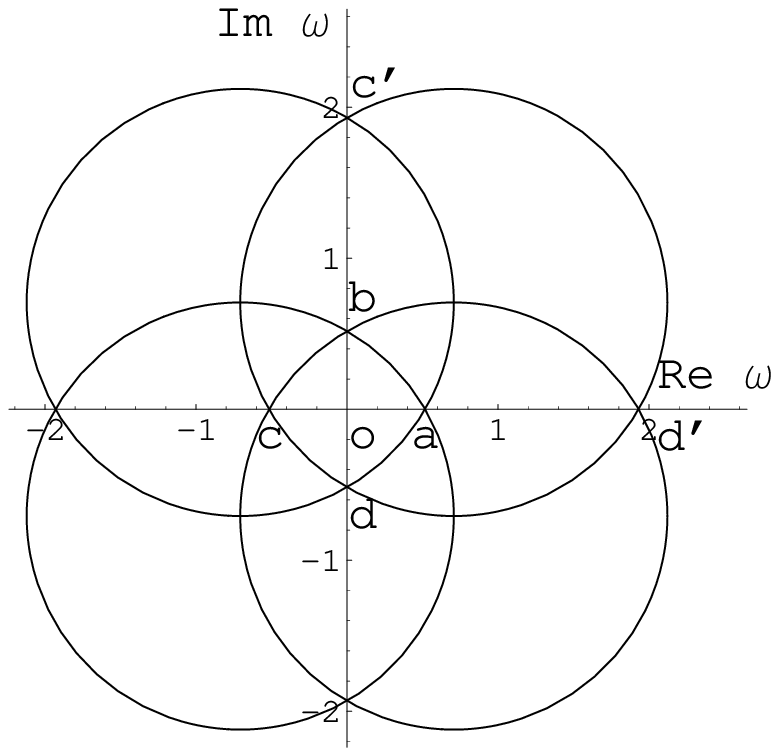}
\caption{Domains of the area in the complex $\omega$ plane which are mapped to the Schwarz's minimal surface \cite{Nitche}. }\label{schwarz_omg}
\end{center}
\end{minipage}
\hfill
\begin{minipage}[b]{0.47\linewidth}
\begin{center}
\includegraphics[width=5cm,angle=0,clip]{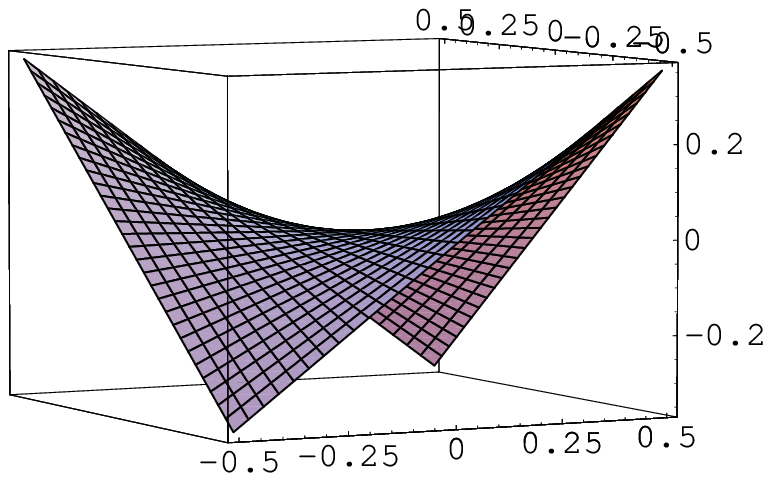}
\caption{A ruled surface with the same boundary as that of Schwarz's minimal surface defined in \cite{Nitche}. }\label{schwarz_ruled}
\end{center}
\end{minipage}
\end{figure}

The domain of the conformal mapping consists of an area bounded by four singular points $a,b,c$ and $d$, where $\displaystyle a=\frac{-1+\sqrt 3}{\sqrt 2}$, $\displaystyle b=\frac{-1+\sqrt 3}{\sqrt 2}i$, $\displaystyle c=\frac{1-\sqrt 3}{\sqrt 2}$ and $\displaystyle d=\frac{1-\sqrt 3}{\sqrt 2}i$\cite{Nitche}. The Schwarz-Christoffel transformation corresponding to the four singular points would be expressed as 
\[
R(\omega)=f(\omega)[(\omega-a)(\omega-b)(\omega-c)(\omega-d)]^{-1/2}.
\] 
The Schwarz reflection principle implies, however, rotation of 180$^\circ$ about the boundary straight line is a symmetry of the mapping and the minimal surface area inside the boundary arc can be reflected to outside the boundary arc.  Taking into account the presence of conjugate singular points, the actual $R(\omega)$ is expressed as
\[
R(\omega)=f(\omega)[(\omega-a)(\omega-b)(\omega-c)(\omega-d)(\omega-a')(\omega-b')(\omega-c')(\omega-d')]^{-1/2},
\]
where $a'=1/b, b'=1/c, c'=1/d$ and $d'=1/a$. The position of the poles in the complex $\omega$ plane are given in Fig.\ref{schwarz_omg}.

We transform $\omega$ to $i\rho$, introduce a scaling parameter $\kappa$ and define 
\[
\displaystyle R(i\rho)=\frac{2i\kappa}{\sqrt{1+14\rho^4+\rho^8}}.
\]
The coordinates of the minimal surface corresponding to the eq.(\ref{schwarz}) scaled by $\kappa$ become
\begin{eqnarray}
\frac{x}{\kappa}= Re \int^\rho \frac{2(1+\rho^2)}{\sqrt{1+14\rho^4+\rho^8}}d\rho\nonumber\\
\frac{y}{\kappa}=-Im \int^\rho \frac{2(1-\rho^2)}{\sqrt{1+14\rho^4+\rho^8}}d\rho\nonumber\\
\frac{z}{\kappa}=Re \int^\rho \frac{4\rho}{\sqrt{1+14\rho^4+\rho^8}}d\rho
\end{eqnarray}
The scaling parameter $\kappa$ is defined at the end of the calculation.

The boundary of the domain of the conformal mapping is bounded by four circles
like
\[
\omega=-\frac{1+i}{\sqrt 2}+\sqrt 2 e^{i\theta}, \quad \frac{\pi}{6}\leq \theta\leq \frac{\pi}{3}.
\]
When $\theta$ varies $\frac{\pi}{6}\to \frac{\pi}{3}$, $\omega$ varies from $\frac{-1+\sqrt 3}{\sqrt 2}\to\frac{-1+\sqrt 3}{\sqrt 2}i$, i.e. $a$ to $b$.

The integral of $x,y,z$ in the Weierstrass-Enneper representation given in sect.2 can be obtained by using the Mathematica,
\begin{eqnarray}
&&\frac{x}{\kappa}=[ 2 \rho  \sqrt{\rho ^4-4 \sqrt{3}+7} \sqrt{\rho
   ^4+4 \sqrt{3}+7}
   \left(F_1\left(\frac{3}{4};\frac{1}{2},\frac{1}{2};
   \frac{7}{4};-\frac{\rho ^4}{7+4
   \sqrt{3}},\frac{\rho ^4}{-7+4 \sqrt{3}}\right) \rho
   ^2\right.\nonumber\\
&&+\left.3
   F_1\left(\frac{1}{4};\frac{1}{2},\frac{1}{2};\frac{
   5}{4};-\frac{\rho ^4}{7+4 \sqrt{3}},\frac{\rho
   ^4}{-7+4 \sqrt{3}}\right)\right)]/{3 \sqrt{\rho
   ^8+14 \rho ^4+1}}
\end{eqnarray}
\begin{eqnarray}
&&\frac{y}{\kappa}=[-2 \rho  \sqrt{\rho ^4-4 \sqrt{3}+7} \sqrt{\rho
   ^4+4 \sqrt{3}+7} \left(\rho ^2
   F_1\left(\frac{3}{4};\frac{1}{2},\frac{1}{2};\frac{
   7}{4};-\frac{\rho ^4}{7+4 \sqrt{3}},\frac{\rho
   ^4}{-7+4 \sqrt{3}}\right)\right.\nonumber\\
&&-\left.3
   F_1\left(\frac{1}{4};\frac{1}{2},\frac{1}{2};\frac{
   5}{4};-\frac{\rho ^4}{7+4 \sqrt{3}},\frac{\rho
   ^4}{-7+4 \sqrt{3}}\right)\right)]/{3 \sqrt{\rho
   ^8+14 \rho ^4+1}}
\end{eqnarray}
\begin{eqnarray}
&&\frac{z}{\kappa}=-\frac{2 i \sqrt{(\rho ^4+4 \sqrt{3}+7)(\rho^4-4 \sqrt 3+7)}
    F\left(i
   \sinh ^{-1}\left(\frac{\rho ^2}{\sqrt{7+4
   \sqrt{3}}}\right)|\frac{7+4 \sqrt{3}}{7-4
   \sqrt{3}}\right)}{\sqrt{(\rho ^8+14 \rho ^4+1)(7-4\sqrt 3)}}.\nonumber\\
&&
\end{eqnarray}
where $F_1(a; b_1,b_2;c;x,y)$ is the Appell's 1st hypergeometric function, $F(\phi,m)$ is the elliptic integral of the first kind.

\subsection{The case of mapping inside the circle $\displaystyle \omega=\frac{1-i}{\sqrt 2}+\sqrt 2 e^{i\theta}$}
The boundary of a circle whose center is at $\displaystyle \frac{1-i}{\sqrt 2}$ in the $\omega$ plane ($\omega=re^{i\alpha}$) is given by
\begin{equation}\label{tanalpha}
\frac{1-i}{\sqrt 2}+\sqrt 2(\cos\theta+i \sin\theta)=r(\cos\alpha+i\sin\alpha)
\end{equation}
We consider an area where $\alpha$ satisfies $0\leq \alpha\leq \pi/2$. The equation 
\begin{equation}
\cot\alpha=\frac{1+2\cos\theta}{-1+2\sin\theta}, \quad r=\sqrt{3+2\cos\theta-2\sin\theta}
\end{equation}
gives a solution of $\theta$ as
\begin{equation}\label{th1}
\theta(\cot\alpha)=\pm \cos^{-1}\frac{1}{2}[(-1-\cot\alpha+\frac{\cot^2\alpha+\cot^3\alpha}{1+\cot^2\alpha}\mp\frac{\cot\alpha\sqrt{3-2\cot\alpha+3\cot^2\alpha}}{1+\cot^2\alpha})]\end{equation}

\begin{figure}[htb]
\begin{minipage}[b]{0.47\linewidth}
\begin{center}
\includegraphics[width=5cm,angle=0,clip]{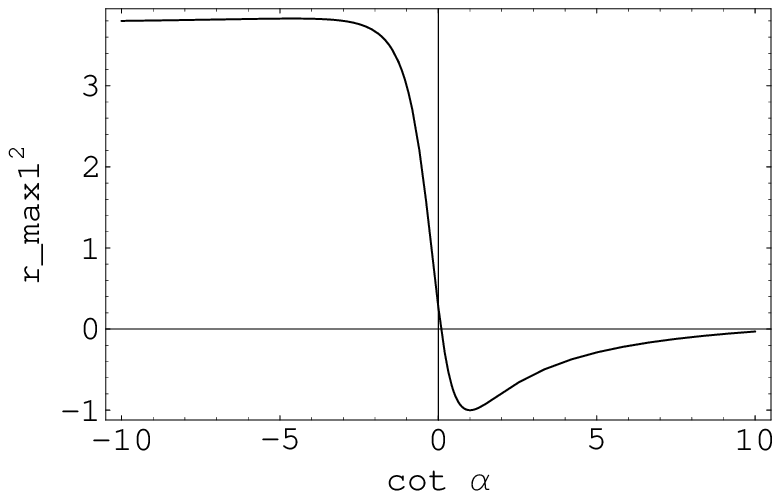}
\caption{The maximal radius $r_{max}$ squared as a function of ${\cot}\alpha$.
In the calculation of the area, the region $0\leq {\cot}\alpha$ is used.}\label{rmax_1}
\end{center}
\end{minipage}
\hfill
\begin{minipage}[b]{0.47\linewidth}
\begin{center}
\includegraphics[width=5cm,angle=0,clip]{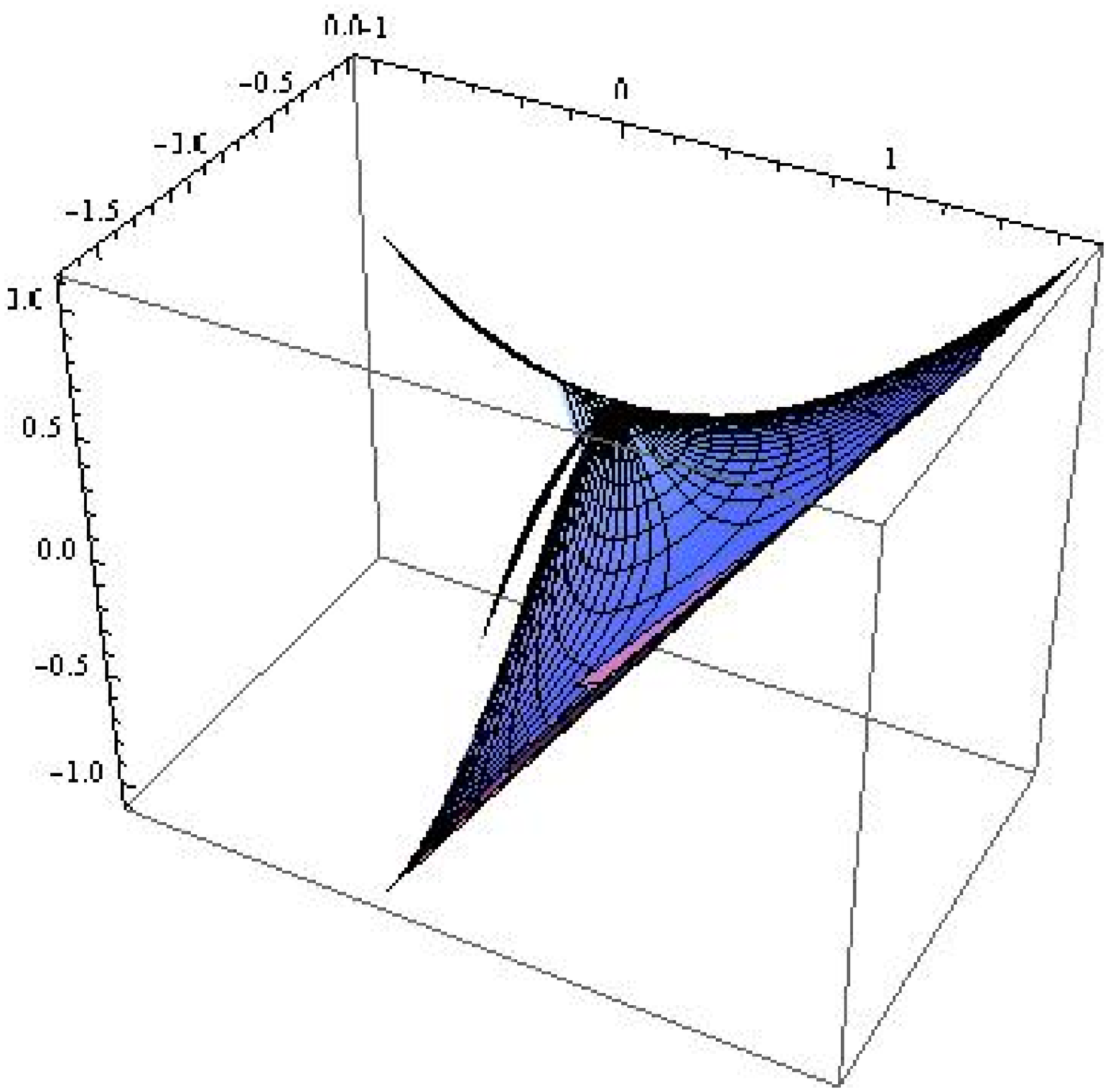}
\caption{A piece of the Schwarz's minimal surface. (The front right piece of the minimal surface of Fig. B.2.)}\label{schwarz1}
\end{center}
\end{minipage}
\end{figure}

Here we choose the first sign + and the second sign - in the eq.(\ref{th1}). Then the $r_{max}^2(\cot\alpha)={3+2\cos[\theta(\cot\alpha)]-2\sin[\theta(\cot\alpha)]}$ as a function of $\cot\alpha$ behaves as Fig.\ref{rmax_1}. There is a branch point at $\cot\alpha=1$, i.e.$\alpha=\pi/4$. 

A mapping of a region $0\leq \theta\leq \pi/2$ and $0\leq r\leq \sqrt{3+2\cos\theta-2\sin\theta}$ via
\begin{eqnarray}
\frac{x}{\kappa}=-Re \int^\rho \frac{2(1+\rho^2)}{\sqrt{1+14\rho^4+\rho^8}}d\rho\nonumber\\
\frac{y}{\kappa}=-Im \int^\rho \frac{2(1-\rho^2)}{\sqrt{1+14\rho^4+\rho^8}}d\rho\nonumber\\
\frac{z}{\kappa}= Re \int^\rho \frac{4\rho}{\sqrt{1+14\rho^4+\rho^8}}d\rho
\end{eqnarray}
is shown in Fig.\ref{schwarz1}.  Due to
the branch point near $\alpha=\pi/4$, there appears numerical errors represented by thorns emanating from the saddle point. The blank area between the thorn going from the saddle point downwards and the left border of the minimal surface is due to numerical difficulties that inhibit simple extension of $\theta$ and $r$ to their boundaries. 

\subsection{The case of mapping inside the circle $\displaystyle \omega=\frac{1-i}{\sqrt 2}+\sqrt 2 e^{i(\theta+\pi/2)}$}

The boundary of the area of a circle whose center is at $\displaystyle\frac{1-i}{\sqrt 2}$ is given by
\begin{equation}
\frac{1-i}{\sqrt 2}+\sqrt 2(\cos(\theta+\pi/2)+i\sin(\theta+\pi/2))=r(\cos\alpha+i\sin\alpha)
\end{equation}

\begin{figure}[htb]
\begin{minipage}[b]{0.47\linewidth}
\begin{center}
\includegraphics[width=5cm,angle=0,clip]{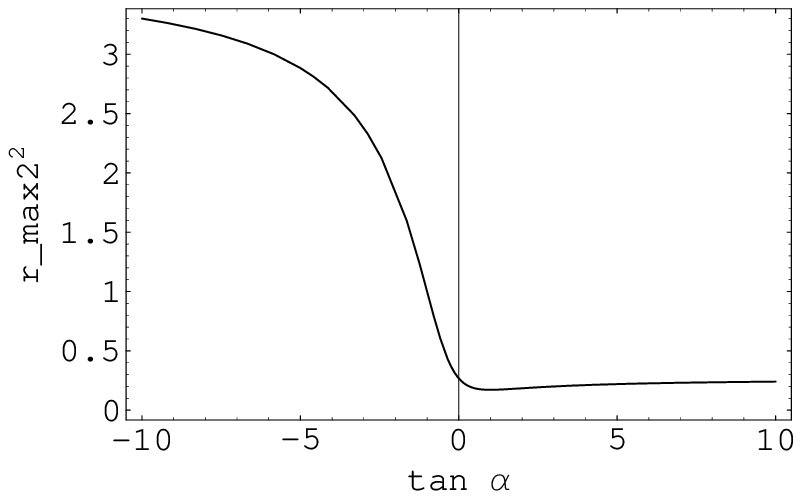}
\caption{The maximal radius $r_{max}$ squared as a function of $\tan \alpha$. In the calulation of the area, the region $\tan\alpha< 0$ is used.}\label{rmax_2}
\end{center}
\end{minipage}
\hfill
\begin{minipage}[b]{0.47\linewidth}
\begin{center}
\includegraphics[width=5cm,angle=0,clip]{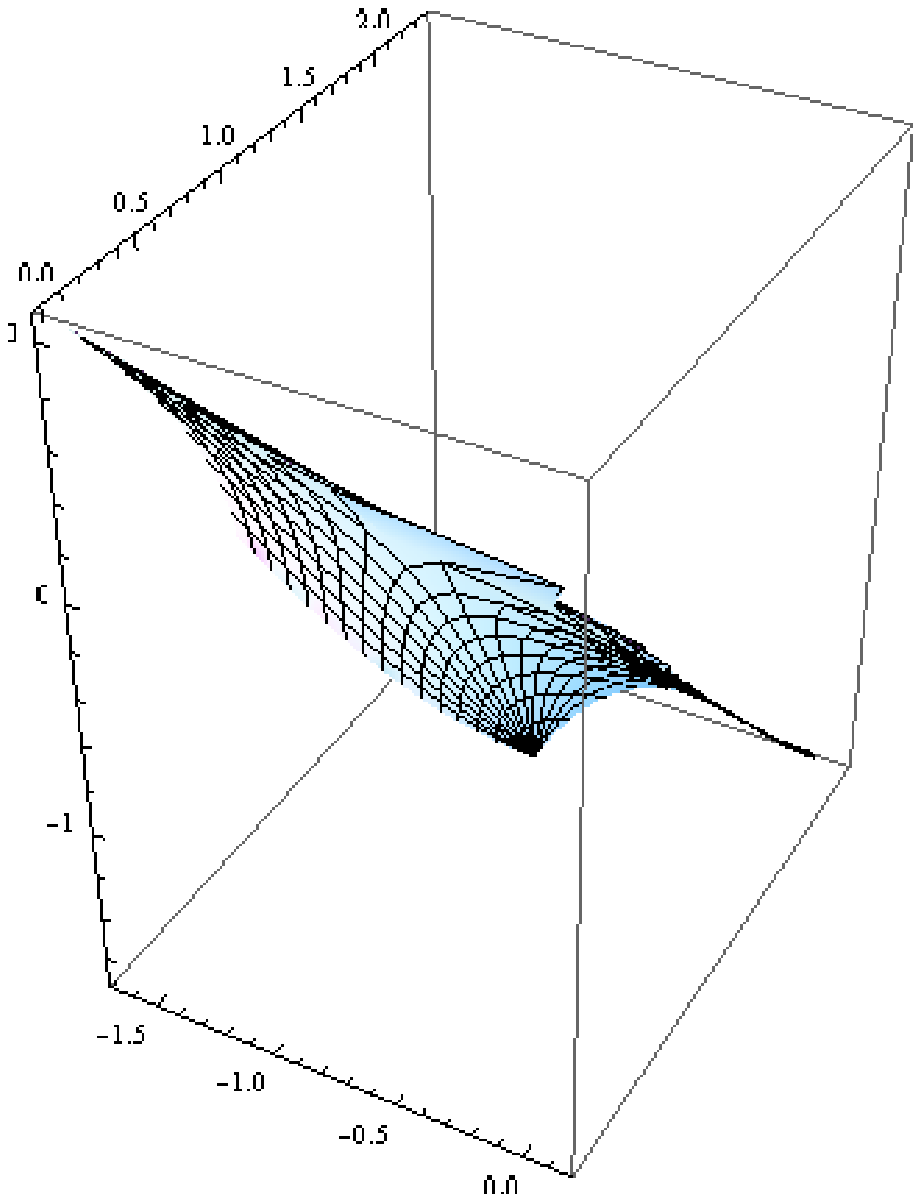}
\caption{A piece of the Schwarz's minimal surface (The front left piece of the minimal surface of Fig. B.2. )}\label{schwarz2}
\end{center}
\end{minipage}
\end{figure}

The equation
\begin{equation}
\tan\alpha=\frac{1+2\cos(\theta+\pi/2)}{-1+\sin(\theta+\pi/2)}=\frac{1-2\sin\theta}{-1+2\cos\theta}, r=\sqrt{3-2\cos\theta-2\sin\theta}
\end{equation}
gives a solution of $\theta$ as
\begin{equation}\label{th2}
\theta(\tan\alpha)=\pm \cos^{-1}\frac{1}{2}[(1-\tan\alpha-\frac{\tan^2\alpha-\tan^3\alpha}{1+\tan^2\alpha}\mp\frac{\tan\alpha\sqrt{3+2\tan\alpha+3\tan^2\alpha}}{1+\tan^2\alpha})]
\end{equation}
Here we choose both the first and the second sign + in the eq.(\ref{th2}). 
The $r_{max}(\tan\alpha)$ squared as a function of $\tan\alpha$ is shown in Fig.\ref{rmax_2}. 

\begin{figure}[htb]
\begin{minipage}[b]{0.47\linewidth}
\begin{center}
\includegraphics[width=5cm,angle=0,clip]{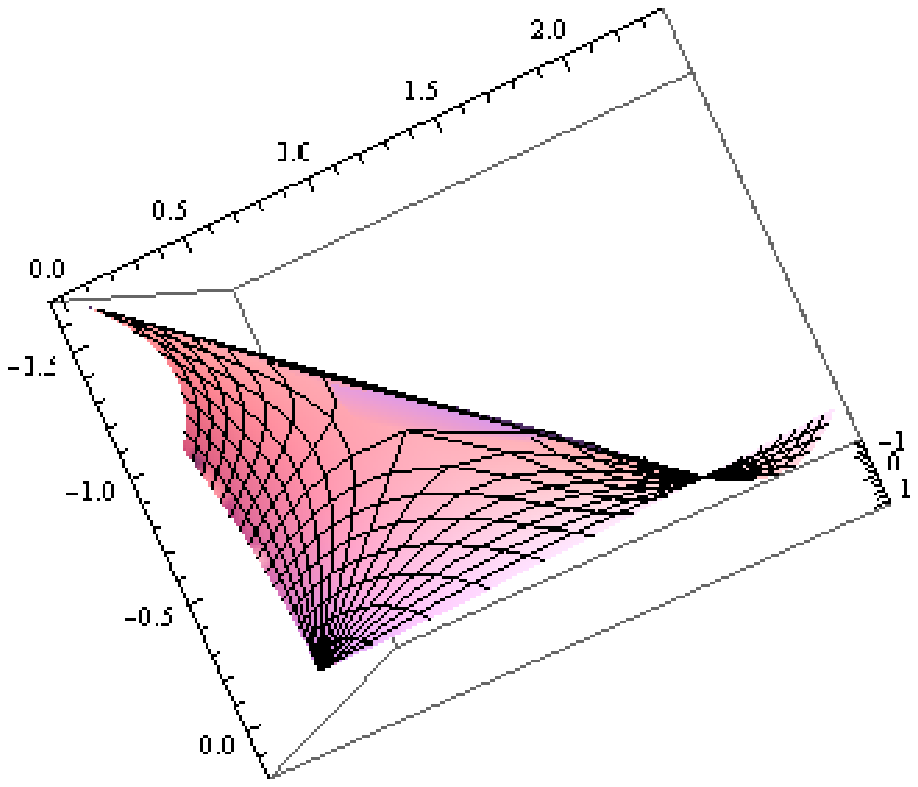}
\caption{The bird's eye view of the minimal surface of Fig.\ref{schwarz2}. The missing corner region in that figure appears in the $90^\circ$ rotated region.}\label{schwarz2up}
\end{center}
\end{minipage}
\hfill
\begin{minipage}[b]{0.47\linewidth}
\begin{center}
\includegraphics[width=5cm,angle=0,clip]{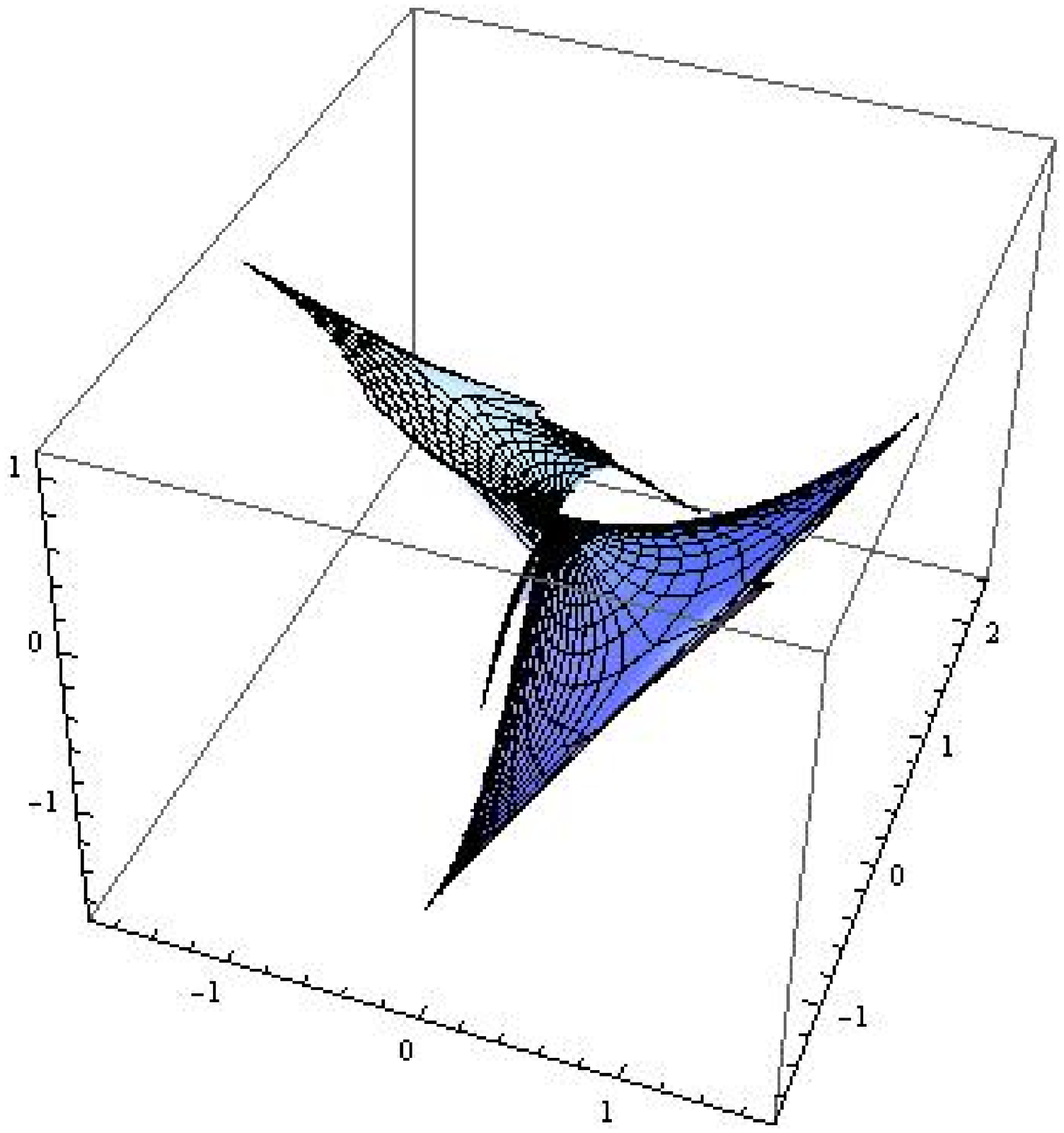}
\caption{A combination of the two pieces of the Schwarz's minimal surface. }\label{schwarz12}
\end{center}
\end{minipage}
\end{figure}

A mapping of a region $-\pi/2 \leq \theta\leq 0$ and $0\leq r\leq \sqrt{3-2\cos\theta-2\sin\theta}$ via
\begin{eqnarray}
\frac{x}{\kappa}=-Re \int^\rho \frac{2(1+\rho^2)}{\sqrt{1+14\rho^4+\rho^8}}d\rho\nonumber\\
\frac{y}{\kappa}=-Im \int^\rho \frac{2(1-\rho^2)}{\sqrt{1+14\rho^4+\rho^8}}d\rho\nonumber\\
\frac{z}{\kappa}= Re \int^\rho \frac{4\rho}{\sqrt{1+14\rho^4+\rho^8}}d\rho
\end{eqnarray}
is shown in Fig.\ref{schwarz2}, which gives a part of the minimal surface. There appears a missing region in upper left corner due to the singularity of $\tan\alpha$ near $\alpha=90^\circ$. The small missing region is mapped in a place rotated by $90^\circ$ and $z<-2\times 0.47196$ region as shown in Fig.\ref{schwarz2up}.

To evaluate the area of the minimal surface, we use data of the right half of the triangle of Fig.\ref{schwarz2up}. 

A combination of the Figs.\ref{schwarz1} and \ref{schwarz2} are shown in Fig.\ref{schwarz12}. The scale parameter $\kappa$ is defined by the height of the $z-$coordinate of the edge of the tetrahedron
\begin{eqnarray}
&&\int_0^{(\sqrt 3-1)i/\sqrt 2}\frac{4\rho}{\sqrt{1+14\rho^4+\rho^8}}d\rho\nonumber\\
&&=-2 i \left(2+\sqrt{3}\right) F\left(-i \sinh ^{-1}\left(7-4
   \sqrt{3}\right)|97+56 \sqrt{3}\right)=-0.47196
\end{eqnarray}
The Fig.\ref{schwarz12} indicates that the actual height of the $z-$ coordinate is twice of this value and since this height should be $\displaystyle \frac{1}{2\sqrt 2}$, $\displaystyle \kappa=\frac{1}{2\sqrt 2}/(2\times 0.47196)=0.37456$.

We observe that the scale factor given in Ref.\cite{Nitche} does not agree with ours.  
To the best of our knowledge, it is the first explicit calculation of the exact minimal surface whose bounding contour is the skew quadrilateral.


\begin{thebibliography}{99}
\bibitem{docarmo} 
M. Do Carmo, \textit{Differential Geometry of Curves and
Surfaces}, Prentice Hall, 1976.
\bibitem{Courant} R. Courant, \textit{Dirichlet's Principle, Conformal Mapping, and Minimal Surfaces}, Springer, New York, Reprint (1977).
\bibitem{farin} 
G. Farin, \emph{Curves and Surfaces for Computer Aided
Geometric Design} 4th ed., Academic Press, 1996.
\bibitem{Nitche} J.C.C. Nitche, \textit{Lectures on Minimal Surfaces}, Cambridge University Press, 1989.
\bibitem{Osserman} R. Osserman, \textit{A Survey of Minimal Surfaces}, Dover Phoenix Editions, Mineola, New York, 1969.
\bibitem{NR} 
W.H. Press et al., in \textit{Numerical Recipes in C++},
Cambridge Univ. Press, 2002
\bibitem{Schwarz} H.A. Schwarz, \textit{Gesammelte mathematische Abhandlungen}, 2 vols, Springer, Berlin, 1890.
\bibitem{businger} 
W. Businger, P.-A. Chevalier, N. Droux and W. Hett,
Mathematica Journal, \textbf{4}(1994) 70.
\bibitem{dziuk} G. Dziuk, \textit{An algorithm for evolutionary surfaces}, Numer. Math. {\bf 58}(1991) 603.
\bibitem{FGM} S. Furui, A.M. Green and B. Masud, \textit{An Analysis of Four-quark Energies in SU(2) Lattice Monte Carlo based on the Cubic Symmetry}, \NPA{\bf 582} 682 (1995).
\bibitem{Karcher} H. Karcher, \textit{The Triply Periodic Minimal Surfaces of A. Schoen and their Constant Mean Curvature Compagnions}, Man. Math. {\bf 64}, 291 (1989).
\bibitem{wrep} F. J. Lopez and
F. Marin, \textit{Complete minimal surfaces in} $R^{3}$, Publicacions Matem\`{a}tiques, Vol. 43 (1999) 341-449.
\bibitem{pipo} 
U. Pinkall and K. Polthier, \textit{Computing Discrete Minimal Surfaces and their Conjugates}, Experim. Math., \textbf{2}(1993) 15.
\bibitem{math} The software "Mathematica" ver. 5.2 \& 6(Prerelease), developed by the Wolfram research.
\end{thebibliography}
\end{document}